\documentclass[12pt,document,nofootinbib,superscriptaddress, onecolumn,preprintnumbers,balancelastpage]{revtex4-1}
\usepackage{amssymb,amsmath,amsfonts,amstext,graphics, multirow,slashed,latexsym,framed}
\usepackage{graphicx}
\usepackage{epstopdf}
\usepackage{url}
\usepackage{appendix}
\usepackage{color}
\usepackage{comment} 
\usepackage{soul}
\usepackage{slashed}
\usepackage{cancel}
\usepackage{titlesec}
\usepackage{setspace}
\usepackage{xspace}
\usepackage{multirow}
\usepackage[table]{xcolor}
\usepackage{colortbl}
\usepackage{footmisc}
\usepackage[margin=1in]{geometry}

\newdimen\tableauside\tableauside=1.0ex
\newdimen\tableaurule\tableaurule=0.4pt
\newdimen\tableaustep
\def\phantomhrule#1{\hbox{\vbox to0pt{\hrule height\tableaurule width#1\vss}}}
\def\phantomvrule#1{\vbox{\hbox to0pt{\vrule width\tableaurule height#1\hss}}}
\def\sqr{\vbox{%
  \phantomhrule\tableaustep
  \hbox{\phantomvrule\tableaustep\kern\tableaustep\phantomvrule\tableaustep}%
  \hbox{\vbox{\phantomhrule\tableauside}\kern-\tableaurule}}}
\def\squares#1{\hbox{\count0=#1\noindent\loop\sqr
  \advance\count0 by-1 \ifnum\count0>0\repeat}}
\def\tableau#1{\vcenter{\offinterlineskip
  \tableaustep=\tableauside\advance\tableaustep by-\tableaurule
  \kern\normallineskip\hbox
    {\kern\normallineskip\vbox
      {\gettableau#1 0 }%
     \kern\normallineskip\kern\tableaurule}%
  \kern\normallineskip\kern\tableaurule}}
\def\gettableau#1 {\ifnum#1=0\let\next=\null\else
  \squares{#1}\let\next=\gettableau\fi\next}

\expandafter\def\expandafter\normalsize\expandafter{%
    \normalsize
    \setlength\abovedisplayskip{8pt}
    \setlength\belowdisplayskip{8pt}
    \setlength\abovedisplayshortskip{8pt}
    \setlength\belowdisplayshortskip{8pt}
}

\titleformat{\section}{\center\normalfont\fontsize{14}{15}\bfseries}{\thesection.}{1em}{}
\titleformat{\subsubsection}{\center\normalfont\fontsize{12}{15}}{\thesubsubsection.}{1em}{}

\definecolor{colorLink}{rgb}{0.7,0,0}
\definecolor{colorCite}{rgb}{0,.7,0}
\definecolor{colorURL}{rgb}{0,0,0.7}
\usepackage[colorlinks=true,linktocpage=true,linkcolor=black,citecolor=colorCite,urlcolor=colorURL]{hyperref}

\newcommand{\gsim}{\lower.7ex\hbox{$\;\stackrel{\textstyle>}{\sim}\;$}}
\newcommand{\lsim}{\lower.7ex\hbox{$\;\stackrel{\textstyle<}{\sim}\;$}}

\newcommand{\be}{\begin{equation}}
\newcommand{\ee}{\end{equation}}
\newcommand{\beq}{\begin{equation}}
\newcommand{\eeq}{\end{equation}}
\newcommand{\bea}{\begin{eqnarray}}
\newcommand{\eea}{\end{eqnarray}}

\newcommand{\bef}{\begin{figure}[htbp]\begin{center}}
\newcommand{\eef}{\end{center}\end{figure}}

\newcommand{\vev}[1]{\langle #1 \rangle}

\newcommand{\eref}[1]{Eq.~(\ref{#1})}

\newcommand{\SB}{$\leavevmode\cancel{\mathrm{SUSY}}\,$}
\newcommand\bigstrut{%
  \vrule height.7\baselineskip depth.3\baselineskip width 0pt\relax}

\begin{document}
\onecolumngrid
\begin{flushright}
%\normalsize SLAC-PUB-XXXXX\\
%\vskip 2 pt
%\normalsize RUNHETC-2012-XX
\end{flushright}

\title{
\vspace{50 pt}
 Gauge Mediated Mini-Split
}
 
\author{ Timothy Cohen }
\affiliation{
Institute of Theoretical Science, University of Oregon, Eugene, OR 97403  \vskip 4 pt  }

\author{ Nathaniel Craig  }
\affiliation{
Department of Physics, University of California, Santa Barbara, CA 93106 \vskip 4 pt } 

\author{ Simon Knapen }
\affiliation{Berkeley Center for Theoretical Physics, \\[-2pt]
University of California, Berkeley, CA 94720 \vskip 3 pt}
\affiliation{Theoretical Physics Group, \\[-2pt]
Lawrence Berkeley National Laboratory, Berkeley, CA 94720 \vskip 4 pt  }

\begin{abstract}
\vskip 20 pt
\begin{center}
{\bf Abstract}
\end{center}
\vskip -6 pt
$\quad$ 
We propose a simple model of split supersymmetry from gauge mediation.  This model features gauginos that are parametrically a loop factor lighter than scalars, accommodates a Higgs boson mass of 125 GeV, and incorporates a simple solution to the $\mu-b_\mu$ problem.  The gaugino mass suppression can be understood as resulting from collective symmetry breaking.  Imposing collider bounds on $\mu$ and requiring viable electroweak symmetry breaking implies small $a$-terms and small $\tan \beta$ -- the stop mass ranges from $10^5$ to $10^8 \mbox{ GeV}$.   In contrast with models with anomaly + gravity mediation (which also predict a one-loop loop suppression for gaugino masses), our gauge mediated scenario predicts aligned squark masses and a gravitino LSP.  Gluinos, electroweakinos and Higgsinos can be accessible at the LHC and/or future colliders for a wide region of the allowed parameter space.
\end{abstract}

\maketitle
\newpage

\section{Introduction}
The discovery of a Standard Model-like Higgs boson at 125 GeV marks the completion of the Standard Model (SM) \cite{:2012gu, :2012gk}.  At the same time, the allowed parameter space for ``natural" supersymmetric scenarios with stops beneath a TeV and decoupled first- and second-generation squarks \cite{Dimopoulos:1995mi, Cohen:1996vb} has been steadily shrinking
~\cite{ATLAS-CONF-2015-066, ATLAS-CONF-2015-067, ATLAS-CONF-2015-078, CMS-PAS-SUS-15-007, CMS-PAS-SUS-15-004, CMS-PAS-SUS-15-005, CMS-PAS-SUS-15-003, CMS-PAS-SUS-15-002}.
%Now super lame 8 TeV results: \cite{Aad:2015pfx,CMS-PAS-SUS-14-011,CMS-PAS-SUS-13-009,CMS-PAS-SUS-13-015,Chatrchyan:2013xna, }
If naturalness is not the correct guiding principle for physics beyond the Standard Model, the time is ripe to study other motivated, testable scenarios that predict new physics at the TeV scale.
 
Supersymmetric extensions of the Standard Model remain strongly motivated in the unnatural limit. Supersymmetry is still one of the few known frameworks that renders the Higgs mass calculable.  Furthermore, accommodating the observed Higgs mass in the context of the Minimal Supersymmetric Standard Model (MSSM)~\cite{Dimopoulos:1981yj} imposes an upper limit on stop mass of $ \lesssim 10^6$ TeV~\cite{Giudice:2011cg,Vega:2015fna}.  This bounds a scale of new physics in supersymmetric theories. Likewise, the successful prediction of gauge coupling unification within the MSSM~\cite{Dimopoulos:1981yj} still remains one of the most compelling infrared indications for any new physics in the ultraviolet.  The most important parameter for precision unification is the Higgsino mass $\mu$.  Consistency at the $2\,\sigma$ level requires $\mu \lesssim 100 \mbox{ TeV}$ \cite{Arvanitaki:2012ps, ArkaniHamed:2012gw}. These considerations strongly motivate the study of split supersymmetry~\cite{Wells:2004di, ArkaniHamed:2004fb, Giudice:2004tc, ArkaniHamed:2004yi} in the ``mini-split"  parameter range, where scalars are beneath $10^6$ TeV~\cite{Arvanitaki:2012ps, ArkaniHamed:2012gw}. 

Although neither the observed Higgs mass nor the suggestion of gauge coupling unification guarantee that superpartners will be experimentally accessible, there are broad classes of models with light gauginos and higgsinos that give rise to signatures at the LHC and other experiments. The exploration of these models provides a productive strategy for beyond the Standard Model searches near the TeV scale.  Among these models, perhaps the most interesting and predictive are those in which the separation between fermionic and scalar superpartners can be explained dynamically by one (or two) loop factors:
\beq\label{eq:DesiredSpectrum}
M_{\tilde{\lambda}} \sim \frac{1}{16 \,\pi^2}\,m_{\tilde{f}},
\eeq
where $M_{\tilde{\lambda}}$ is a gaugino mass and $m_{\tilde{f}}$ is a sfermion mass. Such a one-loop hierarchy occurs in a number of examples.  The most popular of these is anomaly mediation~\cite{Randall:1998uk, Giudice:1998xp} + gravity mediation, where the scalar soft masses are un-sequestered~\cite{Ibe:2012hu, Bhattacherjee:2012ed, Kane:2012qr, ArkaniHamed:2012gw}. These models have the virtue of considerable simplicity and predictivity (for other non-gravity mediation split supersymmetry models, see~\cite{Seiberg:2008qj, Elvang:2009gk, Baryakhtar:2013wy, Kahn:2013pfa, Hook:2015tra}). In this context an observation of the gluino and/or electroweakinos at the LHC would imply that the MSSM scalars are $\sim 100$ TeV -- comfortably within the range suggested by the Higgs mass. Additionally, this mass scale is low enough that indirect evidence for sfermions may appear in the form of experimentally accessible dimension-5 and dimension-6 operators, flavor-conserving or otherwise.  

In practice, putting scalars a loop factor above the TeV scale does not itself adequately insulate the Standard Model from flavor-violating effects due to the squark sector~\cite{Altmannshofer:2013lfa}. Furthermore, meson oscillations \cite{Altmannshofer:2013lfa}, lepton flavor-violating transitions \cite{Altmannshofer:2013lfa}, and (in the presence of new CP-violating phases in the squark mass matrices) electric dipole moment bounds~\cite{McKeen:2013dma, Altmannshofer:2013lfa} are in considerable tension with squarks at the 100 TeV scale. This poses a problem for models with un-sequestered anomaly mediation, which predict $\mathcal{O}(100 \mbox{ TeV})$ anarchic squark masses.
 
In light of these considerations, it is particularly worthwhile to consider mini-split models involving gauge mediation where flavor blindness is generically accommodated. It is well known that gaugino masses can be suppressed in non-minimal models of gauge mediation.  This is clear in the framework of General Gauge Mediation~\cite{Meade:2008wd, Buican:2008ws} where the sfermion and gaugino masses result from independent parameters.  Light gauginos can arise for a variety of reasons~\cite{Nelson:1993nf, Komargodski:2009jf, ArkaniHamed:1998kj, Cohen:2011aa};  unsuppressed gaugino masses are more the exception than the rule. Note also that the particular spectrum in \eref{eq:DesiredSpectrum}\footnote{Many theories with $F/M^2$ suppression of the leading one-loop gaugino masses accumulate $F/M^2$-unsuppressed contributions at three loops; if this is the dominant contribution, the gaugino masses are separated from the sfermions by two loops. This is at the edge of the comfortable range for the Higgs mass, but such models, \emph{e.g.}~\cite{Seiberg:2008qj, Elvang:2009gk, Kahn:2013pfa}, would be interesting to study in their own right.} was achieved in the context of Yukawa-gauge mediation \cite{Galli:2012jp} and semi-direct gauge mediation with chiral messengers~\cite{Argurio:2009ge, Argurio:2010fn}. In the former model, the gaugino and sfermion masses may arise at two and three loops respectively. The latter setup requires an extra gauge group, implying that it is non-trivial to generate a suitable Higgs potential.  Therefore, it remains interesting to discover alternative models which explain this one-loop suppression as a consequence of a symmetry argument. 

There is another motivation for exploring gauge-mediated mini-split supersymmetric theories. This is the ``$\mu$ - $b_\mu$ preference," the tendency for calculable models of the Higgs sector parameters to generate a $b_\mu$ soft term that is a loop factor larger than the corresponding $\mu$ term squared.  Usually, $b_\mu \sim m_{\tilde{f}}^2$ for viable electroweak symmetry breaking, suggesting that the Higgsino mass is typically a loop factor lighter than $m_{\tilde{f}}$.  While this is a problem for theories with weak-scale sfermions on account of collider limits, in mini-split theories the entire spectrum shifts to higher scales and the $\mu$ - $b_\mu$ ``problem'' is recast as a preference for parametrically light Higgsinos. Heavy sfermions also resolve the challenge of explaining the Higgs mass in gauge-mediated models \cite{Draper:2011aa,Knapen:2015qba}. Theories combining the gauge-mediated preference for loop-suppressed Higgsinos with gaugino masses satisfying \eref{eq:DesiredSpectrum} are exceptionally predictive targets for the LHC and/or future colliders.

Taking these various motivations (gauge coupling unification, a 125 GeV Higgs mass, current flavor bounds, the $\mu$ - $b_\mu$ preference, and a desire for observable particles at colliders) together, in this work we construct flavor-blind models that yield the loop factor relationship between $m_{\tilde{f}}$ and $M_{\tilde{\lambda}}$. The $\mu$ - $b_\mu$ preference implies that Higgsinos can be in the same mass range as gauginos.  For the sake of specificity, we will include couplings between the messengers and Higgs doublets to generate Higgs soft parameters as in \cite{Dvali:1996cu, Csaki:2008sr, DeSimone:2011va}.  The ``challenges" of gauge mediation -- suppressed gaugino masses and achieving electroweak symmetry breaking -- become features of gauge mediated mini-split.

The rest of this paper is organized as follows:  In \S\,\ref{sec:ConcreteModel} we propose a concrete model to realize a predictive loop-factor separation between scalar and gaugino masses in gauge mediation.  In \S\,\ref{sec:ParamSpace} we demonstrate the viability of the model and elucidate its characteristic features through a numerical study of the spectrum.  We discuss generic LHC signatures of the model and appropriate search strategies in \S\,\ref{sec:ColliderPheno}.  Finally, in \S\,\ref{sec:Conclusions} we conclude.  

\section{The Model}\label{sec:ConcreteModel}
Our goal is to mediate supersymmetry breaking (\SB\!) to the MSSM via the $SU(3)\times SU(2) \times U(1)$ SM gauge interactions such that a loop factor hierarchy between the soft masses for the gauginos and the scalars is realized dynamically.  Schematically, this is achieved by charging the source of \SB\!, a spurion superfield $X$ with $\vev{X}=F \,\theta^2 $, under one or more spurious symmetries which forbid the gaugino mass operator
\begin{equation}\label{gauginoOp}
\mathcal{L}\supset \int\!\!d^2\theta\; \frac{X}{M} \,\mathcal{W}^\alpha\, \mathcal{W}_\alpha
\end{equation}
where $M$ parametrizes the typical scale of the messengers. If these spurious symmetries are then broken explicitly by one or more marginal operators, the gaugino mass term in \eref{gauginoOp} will be generated, but only at higher loop order. Conceptually this setup is identical to the idea of collective symmetry breaking in little Higgs models \cite{ArkaniHamed:2001nc}, see for instance \cite{Schmaltz:2005ky} and references therein.   The following subsection will exemplify the collective breaking mechanism by explicitly computing the gaugino and sfermion masses.  Once we have determined the messenger scale values of the squark, slepton, and gaugino masses, we will move on to the details of the Higgs sector in \S\,\ref{sec:mubmu}.  This provides a simple and calculable realization of the $\mu$ - $b_\mu$ preference \cite{Dvali:1996cu,Csaki:2008sr, DeSimone:2011va}.

\subsection{Gaugino and sfermion masses}

Consider the following model of messenger-spurion interactions, given by a superpotential
 \begin{align}\label{model}
 W \supset &\, \big( X + \zeta_2\, S \big)\, \bar{\phi}_1\,\phi_2+ M_\phi\, \big(\,\bar{\phi}_1\,\phi_1+\bar{\phi}_2\,\phi_2\big), 
 \end{align}
where the $\phi\;\big(\bar\phi\big)$ are ${\bf 5}$ $\big({\bf \bar 5}\big)$ messenger fields, and $S$ a gauge singlet pseudo-modulus, and $\zeta_2$ is a spurion for breaking of a global symmetry $G_2$ that is introduced below. Aside from the gauge symmetry, the non-zero couplings are determined by the trivial $R$-symmetry and a global symmetry $G_1$ with charge assignments 
\begin{equation}\label{sym1}
G_1\big[\bar{\phi}_1\big]=-G_1\big[\phi_1\big]=-G_1\big[X\big]=-G_1\big[S\big]=1 \quad \mathrm{and}\quad G_1\big[\phi_2\big]=G_1\big[\bar \phi_2\big]=0.
\end{equation}
The gaugino masses are not generated at any order in $F/M_\phi^2$ as a consequence of $G_1$ and the $R$-symmetry.  Using standard methods~\cite{Cheung:2007es, Dumitrescu:2010ha}, the soft spectrum resulting from this messenger sector can be computed:
\begin{align}
M_i&=0;\label{gauginowrong}\\
m_{\tilde f}^2&= 2\sum_{i=1}^3\left(\frac{\alpha_i}{4\,\pi}\right)^2 C^i_{\tilde f} \frac{F^2}{M_\phi^2},
\end{align}
where the index $i$ labels the gauge group and the $C^i_{\tilde f}$ are the appropriate quadratic Casimirs. Here we assume $F\ll M_\phi^2$ such that $F/M_\phi^2$ suppressed contributions to the scalar masses are neglected.

Integrating out the $\phi$ messengers generates a potential for the pseudo-modulus $S$, yielding the effective K\"ahler potential
\begin{align}
\mathcal{K}_\text{eff}\supset-\frac{1}{32\,\pi^2}\Bigg[\zeta_2\, S\,X^\dagger \left(2+\log\frac{M_\phi^2 }{\Lambda^2}\right)+\mathrm{h.c.} +\frac{2}{3}\,\big|\zeta_2\big|^2 \frac{\big|X\big|^2}{M_\phi^2} \big|S\big|^2\Bigg],
\end{align} 
which stabilizes $S$ at the origin of moduli space. In addition, a vacuum expectation for its $F$-component is induced:
\begin{equation}\label{eq:Fs}
F_S= \frac{\zeta_2^\dagger}{16\,\pi^2} F,
\end{equation} 
where we have chosen the scheme corresponding to setting $\Lambda=M_\phi$.

While this model does yield suppressed gaugino masses, it does not yet achieve the spectrum in \eref{eq:DesiredSpectrum}. This can be resolved by coupling another set of ${\bf5}$ - ${\bf \bar 5}$ messengers to $S$ as follows 
\begin{equation}\label{eq:Wextra}
W\supset\big(\zeta_1\, S+ M_\chi \big)\bar\chi\,\chi.
\end{equation}
Upon integrating out the $\chi$, the effective superpotential contains the term
\begin{equation}
W_\text{eff}\supset \zeta_1 \,\frac{\alpha_i}{4\,\pi}\frac{S}{M_\chi} W_i^\alpha \,W_{i,\alpha}\,,
\end{equation}
where $\zeta_1$ is a spurion for $G_1$ symmetry breaking.  In combination with \eref{eq:Fs}, this generates non-zero gaugino masses
\begin{equation}
M_i=\frac{\zeta_1\,\zeta_2^\dagger}{16\,\pi^2}  \frac{\alpha_i}{4\,\pi}\frac{F}{M_\chi},
\end{equation}
yielding the desired relation if $M_\chi\sim M_\phi$. Note that \eref{eq:Wextra} will also contribute to the sfermion masses-squared, but only at four loops; this contribution is neglected in what follows.

The features of these results can be understood easily in terms of symmetries. From \eref{eq:Wextra} one can see that $\zeta_1$ is a symmetry breaking spurion for $G_1$. In addition, in the limit where $\zeta_2=0$, the model posses a second global symmetry $G_2$ with charge assignments 
\begin{equation}\label{sym2} 
G_2\big[\bar\phi_1\big]=-G_2\big[\phi_1\big]=-G_2\big[X\big]=1 \quad \mathrm{and}\quad G_2\big[\phi_2\big]=G_2\big[\bar \phi_2\big]=G_2\big[S\big]=0,
\end{equation}
which would also forbid the $F/M_\phi$ contribution to the $M_i$. The gaugino masses can only be generated if $G_1$ and $G_2$ are broken \emph{collectively} by the presence of both the spurions $\zeta_1$ and $\zeta_2$. This in turn implies the presence of the additional loop factor.  

Note that a vanishing (or somewhat suppressed) lowest component for $X$ has been assumed. If $\langle X \rangle \neq 0$, the symmetries protecting the gauginos are broken by a relevant operator and the gaugino masses would again appear at one loop:
\begin{equation}\label{eq:gauginovictory}
M_i \sim\frac{\alpha_i}{4\,\pi} \frac{\langle X \rangle }{M_\phi}\frac{F^3}{M^5_\phi}.
\end{equation}
This expression reduces to that of (extra)ordinary gauge mediation with the appropriate choice of $R$-charges if $\langle X \rangle=M_\phi$~\cite{Cheung:2007es} (See also \cite{Nomura:1997uu}). Finally note that the conventional wisdom regarding gaugino screening does not apply here, since $S$ is a pseudo-modulus rather than a heavy messenger \cite{Dumitrescu:2010ha,ArkaniHamed:1998kj,Cohen:2011aa}.  A similar exception to the gaugino screening theorem exists for chiral messengers \cite{Argurio:2010fn}.

%%%%%%%%%%%%%%%%%%%%%%%%%%%%%%%%%%%%%%%
\subsection{$\mu$ and $b_\mu$}\label{sec:mubmu}

The LHC constraints on the gluino of $M_3 \gtrsim 1.5 \mbox{ TeV}$ translate into a prediction for the sfermion masses $m_{\tilde f} \gtrsim \mathcal{O}(100 \mbox{ TeV})$.  In order to reproduce the Higgs boson mass, we are naively pushed to $\tan \beta \lesssim 5$ \cite{Giudice:2011cg, Vega:2015fna}.  (The more detailed analysis shown in Fig.~\ref{fig:ParamSpace} below demonstrates that the LEP bound on $\mu$ and the BBN bound on the gravitino provide the most stringent lower limits $\tan \beta$, but the qualitative story is unchanged.)  Such small values of $\tan \beta$ are compatible with electroweak symmetry breaking driven by the $b_\mu$ term, with the $\mu$ term playing little role. This is readily compatible with $ b_\mu \gg \mu^2$, which is a generic feature in many simple gauge-mediated models that generate the Higgs soft parameters \cite{Dvali:1996cu, Csaki:2008sr, DeSimone:2011va}. As such, there is no $\mu$ - $b_\mu$ problem in this setup.

We can gain some insight into the implications of  $\mu^2 \ll b_\mu$ using the tree-level electroweak symmetry breaking (EWSB) conditions.\footnote{We will include the 1-loop contribution to the effective potential from the top and stops when computing example spectra in \S\,\ref{sec:ParamSpace} below.  These contributions are non-trivial in models of split supersymmetry, but the tree-level EWSB conditions still provide a useful qualitative guide.}  In this region of parameter space the MSSM vacuum stability conditions become
\be\label{eq:bmuForGoodEWSB}
2\, b_\mu \lesssim {m}_{H_u}^2 + {m}_{H_d}^2; \quad\quad\quad b_\mu^2 \gtrsim {m}_{H_u}^2\, {m}_{H_d}^2,
\ee
where $b_\mu$ is taken to be real and positive, and ${m}_{H_{u,d}}$ are the Higgs soft-mass squared parameters.
The only way to satisfy both of these conditions is for ${m}_{H_{u,d}}^2 \gtrsim 0$.  This restriction is important for understanding the details of the parameter space for the Higgs and stop soft masses \cite{Arvanitaki:2012ps}.

Next we introduce simple Higgs-messenger couplings, such as those found in the ``lopsided" model of \cite{Csaki:2008sr, DeSimone:2011va}, in order to realize a viable (and calculable) Higgs sector.\footnote{This model may also yield a solution to the $\mu$ - $b_\mu$ for non-split spectra if the hidden sector is strongly coupled \cite{Knapen:2013zla}.}  We couple the Higgs doublets $H_{u,d}$ to another set of messengers $\chi$ and an additional heavy gauge singlet via the superpotential coupling \cite{Dvali:1996cu,Csaki:2008sr, DeSimone:2011va}
\begin{equation}\label{eq:lopsided}
W\supset \big(X +M_N \big) N\,\bar N + M_\chi\,\chi\,\bar \chi + \lambda_u\, \bar N\, \chi\, H_u + \lambda_d \, N \,\bar\chi \, H_d,
\end{equation}
where it is understood that only the doublet component of $\chi$ ($\bar\chi$) couples to $H_u$ ($H_d$). The full model therefore contains only three  ${\bf5}$ - ${\bf \bar 5}$ multiplets, which is well below the bound from requiring no Landau poles before the GUT scale \cite{Giudice:1998bp}. Additionally, the model is automatically free from large CP violation since no new physical phases are introduced.  Integrating out the messengers results in the following threshold corrections to the soft parameters \cite{DeSimone:2011va}
\begin{align}
\mu =\frac{\lambda_{u}\,\lambda_{d}}{16\,\pi^2} \frac{F}{M_N}\frac{x \left(x^2-\log \left(x^2\right)-1\right)}{\left(x^2-1\right)^2} \,\,\,\quad\quad\quad\quad\quad\quad\quad\quad&\xrightarrow[\,\, x \rightarrow 1 \,\,]{}\quad \frac{\lambda_u\, \lambda_d}{32 \, \pi^2} \frac{F}{M_N};\label{eq:mu}\\
b_\mu =\frac{\lambda_{u}\,\lambda_{d}}{16\,\pi^2}  \left(\frac{F}{M_N}\right)^2\frac{x \left(-x^4+2 x^2 \log \left(x^2\right)+1\right)}{\left(x^2-1\right)^3}\quad\quad\quad \quad&\xrightarrow[\,\, x \rightarrow 1 \,\,]{}\quad-  \frac{\lambda_u \,\lambda_d}{48\, \pi^2} \frac{F^2}{M_N^2};\label{eq:bmu}\\
\delta {m}_{H_{u,d}}^2 =\frac{\lambda_{u,d}^2}{16\,\pi^2} \left(\frac{F}{M_N}\right)^2 \frac{x^2 \left(-2 x^2+\left(x^2+1\right) \log \left(x^2\right)+2\right)}{\left(x^2-1\right)^3} \quad&\xrightarrow[\,\, x \rightarrow 1 \,\,]{}\quad\frac{\lambda^2_{u,d}}{96\, \pi^2} \frac{F^2}{M_N^2};\label{eq:mHsq}\\
 a_{u,d} =\frac{\lambda_{u,d}^2}{16\,\pi^2} \frac{F}{M_N} \frac{ x^2 \left(-x^2+\log \left(x^2\right)+1\right)}{\left(x^2-1\right)^2} \,\,\,\quad\quad\quad\quad\quad\quad\quad&\xrightarrow[\,\, x \rightarrow 1 \,\,]{}\quad - \frac{\lambda^2_{u,d}}{32 \pi^2} \frac{F}{M_N},\label{eq:aTerm}
\end{align}
where $x \equiv M_N/M_\chi$, $\delta {m}_{H_{u,d}}^2$ is an additive contribution to $m^2_{H_{u,d}}$, $a_{u,d}$ are $a$-terms involving the up and down type scalars respectively, and the approximation that $\chi$ does not couple to supersymmetry breaking is taken. 
 
The presence of the $M_N$ mass term in \eref{eq:lopsided} breaks the symmetries in \eref{sym1} and \eref{sym2}. Indeed, \eref{eq:lopsided} contributes to the wino and bino masses through two-loop diagrams involving $H_{u,d}$. Any such contribution is necessarily proportional to $\sim \lambda_{u,d}^2$. As will become clear shortly, $|\lambda_{u,d}|^2\ll 1$ in the viable parameter space such that it is generally safe to neglect these contributions provided that $\zeta_1\sim \zeta_2\sim 1$.  

This completes the details of the model.  In the next section we turn to a discussion of the parameter space.  Emphasis is made on the resultant gaugino masses and $\mu$-parameter since these lead to the near-term observable signatures of this scenario.

\section{Results}\label{sec:ParamSpace}
In this section we will discuss the viability of the spectrum.  We will begin with some simple parametric estimates of various constraints, in order to provide a reliable qualitative guide.  We then present a numerical analysis demonstrating the range of allowed masses.  Large regions of parameter space predict gauginos and Higgsinos that are within the reach of the LHC and/or future colliders.  

\subsection{Constraints on the Parameter Space}\label{sec:Constraints}
As described above, we will be satisfying the EWSB conditions in the regime where \eref{eq:bmuForGoodEWSB} holds.  Concretely this implies 
\be\label{eq:ewsbparam}
b_\mu \sim {m}^2_{H_{u,d}}\quad \Longrightarrow \quad \lambda_u\, \lambda_d \sim \frac{g_2^2}{16\,\pi^2}.
\ee
Although $\mu$ and $b_\mu$ were both generated at one loop, with this assumption they are parametrically comparable to the gaugino and sfermion masses, respectively.

Without large gaugino masses, a significant constraint on the parameter space comes from avoiding prohibitive charge- and color-breaking minima. The fields most likely to become tachyonic are those that see the large top Yukawa coupling, namely $H_u, \tilde t_L,$ and $\tilde t_R$. If there is any substantial mass hierarchy between these fields, renormalization group contributions proportional to the top Yukawa drive the smallest of the three masses negative even with a short amount of running since the countervailing contributions from gaugino masses are negligible.  We must therefore avoid any large hierarchy between the soft masses and as a result tend to live close to the UV fixed point of the renormalization group equations (RGE). In other words, a consistent spectrum favors
\be\label{eq:mHeqmT}
\lambda_{u}^2 \sim \frac{g_3^2}{16\,\pi^2},
\ee
where we have assumed the contribution given in \eref{eq:mHsq} dominates the Higgs soft masses. Using Eqs.~(\ref{eq:ewsbparam}) and (\ref{eq:mHeqmT}), we therefore expect $|\lambda_d|/|\lambda_u|\sim g_2/g_3$.  We will verify numerically in \S\,\ref{sec:Spectrum} that ${m}_{H_u} \simeq {m}_{\tilde t_{1,2}}$ to very good approximation and $|\lambda_d| \lesssim |\lambda_u|$ in the viable parameter space.

When \eref{eq:mHeqmT} holds, the high-scale contribution to the $a$-terms given in \eref{eq:aTerm} is also effectively a two-loop contribution:
\be
a_{u,d} \sim - \frac{g_s^2}{(16\,\pi^2)^2}\frac{F}{M} \ll \sqrt{\bigstrut\, m_{\tilde t_1}\,m_{\tilde t_2}}
\ee
and is therefore of little importance for the mass spectrum. Furthermore, since the gaugino masses are small, the $a$-terms will not be regenerated by RG evolution.  To good approximation we can use the results of \cite{Giudice:2011cg,Vega:2015fna} which assume zero $a$-terms for our determination of the Higgs mass. 

There is a final important RGE effect that we need to consider.  Since ${m}_{H_u}^2 \neq {m}_{H_d}^2$ at the messenger scale, there are potentially large contributions to the soft masses from the hypercharge $D$-terms in the RGEs.  Depending on the relationship between ${m}_{H_u}^2$ and ${m}_{H_d}^2$, this contribution can drive either the right-handed selectron soft mass-squared or the left-handed slepton doublet soft mass-squared to negative values. 
All of these constraints will be satisfied for the spectra we will present in the next section. 

\subsection{The Spectrum}\label{sec:Spectrum}
The previous discussion has been largely qualitative. This section will demonstrate that once the tuning required to reproduce the weak scale has been performed, a typical parameter choice will yield (i) stable electroweak symmetry breaking with $v=246\mbox{ GeV}$, (ii) a Standard Model-like Higgs boson near 125 GeV, (iii) gauginos that are accessible at the LHC or future colliders, (iv) $\mu$ in the range of hundreds of GeV to well beyond a TeV, (v) and a gravitino LSP (with mass $m_{3/2}$).    We will ensure $\mu < 100\mbox{ TeV}$ so that the MSSM gauge couplings unify \cite{Arvanitaki:2012ps, ArkaniHamed:2012gw}.

In order to evolve spectra at the scale $M$ down to the weak scale, we use one-loop RGEs for the soft parameters and two-loop RGEs for the gauge couplings~\cite{Machacek:1983tz, Machacek:1983fi}. 
We evaluate the electroweak symmetry breaking conditions at  the geometric mean of the physical stop masses.  Below this scale we decouple the heavy scalars with a step function. The gaugino masses and $\mu$ are evolved to the weak scale using the RGEs appropriate to split supersymmetry \cite{Giudice:2004tc}.   We numerically solve for $\lambda_{u,d}$ by requiring that the minimum of the potential reproduces the correct vacuum for a given choice of $\tan\beta$. For this purpose we include the one-loop Coleman-Weinberg contribution from the top and stops. 

A priori, the free parameters of the model are 
\begin{equation}
F,\; M_\phi\,\; M_\chi,\; M_N,\; \zeta_1,\;\zeta_2,\;\lambda_u, \text{ and } \lambda_d.
\end{equation}
Many choices of these parameters however lead to the same qualitative physics in the IR. In particular, the masses of the gauginos relative to the sfermions are controlled by the combination
\begin{equation}
\label{eq:MiToMf}
\frac{M_i}{m_{\tilde f}} \sim \frac{M_\phi}{M_\chi}\frac{\zeta_1\,\zeta_2}{16\,\pi^2}.
\end{equation}
Since the spectrum is only sensitive to the combination $\zeta_1\,\zeta_2$, we can define $\zeta\equiv\zeta_1=\zeta_2$ without loss of generality. For concreteness, we further choose
\begin{equation}
M\equiv M_\phi= M_\chi= M_N=10^4\times m_s.
\end{equation}
where $m_s=\sqrt{m_{\tilde{Q}_3}m_{\tilde{U}_3}}$ is the geometric mean of the stop soft masses. This ensures that $F/M^2\sim 10^{-2}$, such that the approximations taken in \S\,\ref{sec:ConcreteModel} are justified. Since $\lambda_{u,d}$ are solved for using the EWSB conditions, the remaining independent parameters can be chosen as
\begin{equation}
m_s,\; \tan\beta, \text{ and } \zeta,
\end{equation}
where the coupling $\zeta$ determines the overall mass scale of the gauginos. Since the gauginos are parametrically lighter than the scalars, their masses do not significantly affect the rest of the spectrum for $\zeta\sim1$. $\zeta$ therefore effectively factorizes from the remaining two parameters.

In Table~\ref{tab:ExampleParams} we give two examples of input parameters and the resultant spectrum.  These cases satisfy all of the criteria enumerated above.  For both cases, the gauginos are accessible at the LHC.  In the first case $\mu$ will be important for determining the composition of the neutralino and charginos while in the second case $\mu$ is essentially decoupled.  This will imply different characteristic signatures at colliders as described below in \S\,\ref{sec:ColliderPheno}.

\begin{table}[h!]
\renewcommand{\arraystretch}{1.}
\setlength{\tabcolsep}{6pt}
\renewcommand{\arraystretch}{2.}
\begin{centering}
Input parameters\\[2pt]
\begin{tabular}{|c||c|c|c||c|c|c|c|}
\hline
example&$m_s$ & $\tan\beta$ & $ \zeta$& $F/M$ &$F/M^2$  & $\lambda_u$ & $\lambda_d$ \\
\hline
small $\mu$&$2.3\times 10^5$ & 2.4 & 1&$ 4.2\times 10^7 $& 0.017 & 0.091& -0.022 \\
large $\mu$ &$8.8\times 10^5$& 2. &$1/2$& $1.7\times 10^8 $& 0.017 & 0.086 & -0.028 \\
 \hline
\end{tabular}\\[8pt]
Output spectrum\\[2pt]
\begin{tabular}{|c||c|c|c|c|c|c|c|c|}
\hline
example & $M_1$ & $M_2$ & $M_3$ & $\mu$ & $\sqrt{\bigstrut\,m_{\tilde t_1} m_{\tilde t_2}}$ &  $\sqrt{m_{Q_{1,2}} m_{U_{1,2}}}$ &$m_h$ &${m}_{3/2}$\\
 \hline
small $\mu$ & 380 & 700 & 2040 & -290 &  $2.4\times 10^5$ &$  2.6\times 10^5$& 126 & 0.025\\
large $\mu$ &400 & 770 & 2180& -1380 & $ 8.7\times 10^5 $& $9.9\times 10^5$ & 126 & 0.41\\
\hline
\end{tabular}
\end{centering}
\caption{Example input parameters and the resulting output spectrum for our model. All dimensionful quantities are in units of GeV. In both cases, the gluino should be accessible at the 13 TeV LHC.  We present an example with ``small" (``large") $\mu$ where the LSP would have a large (small) Higgsino component. The Higgs mass was computed with susyHD \cite{Vega:2015fna}. }
\label{tab:ExampleParams}
\end{table}

\begin{figure}[h!] %  figure placement: here, top, bottom, or page
   \centering
   \includegraphics[width=.8\textwidth]{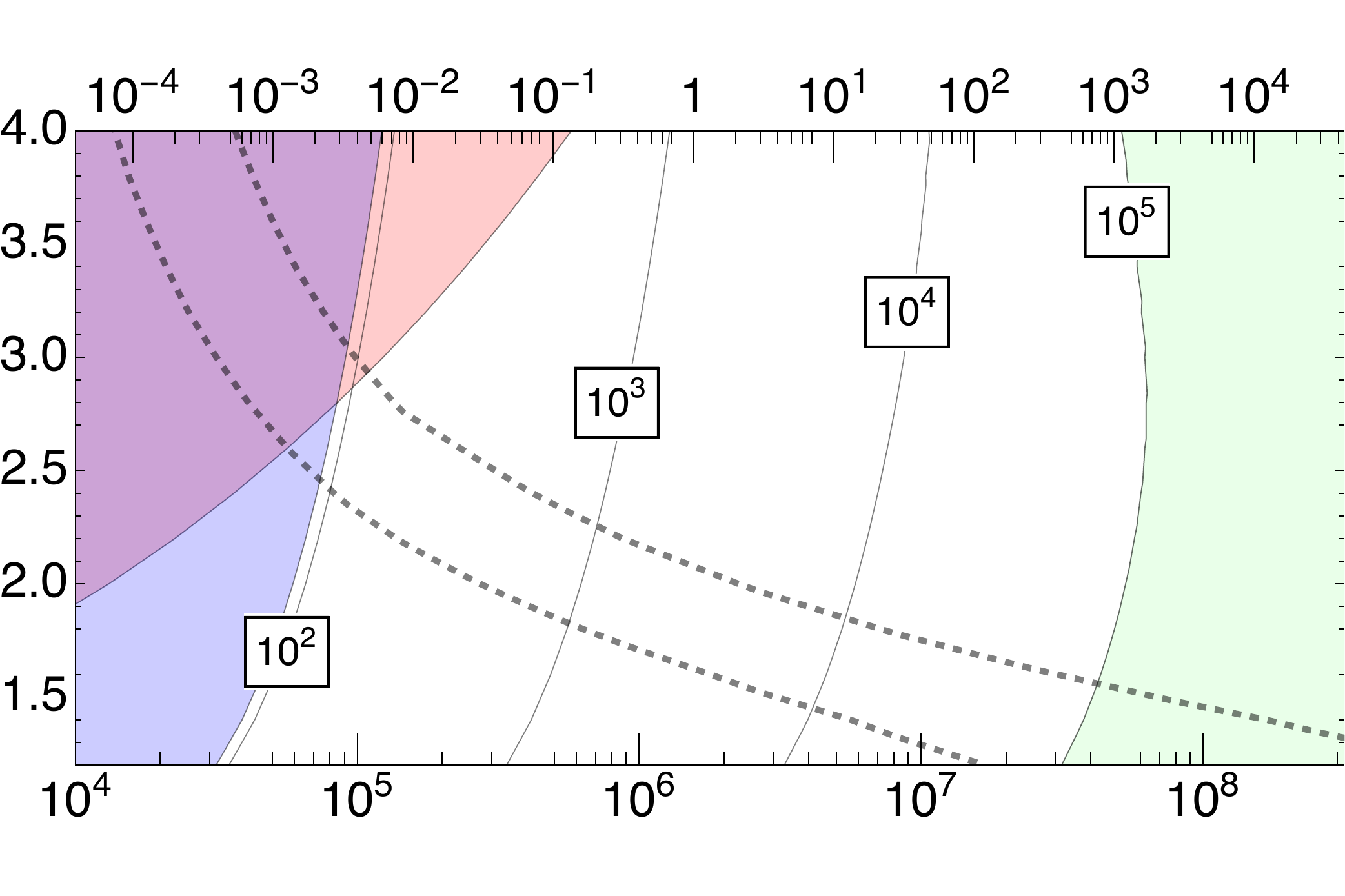}   
   \hspace*{-10mm}
\begin{minipage}{0cm}
\vspace*{.5cm}\hspace*{-10.9cm}{\Large $m_s\,\,\big[\text{GeV} \big]$}
\vspace*{0.5cm}
\end{minipage}
\begin{minipage}{0cm}
\vspace*{-16.6cm}\hspace*{-10.9cm}{\Large $m_{3/2}\,\,\big[\text{GeV} \big]$}
\vspace*{0.5cm}
\end{minipage}
\begin{minipage}{0cm}
\vspace*{-8.8cm}\hspace*{-26.8cm}\rotatebox{90}{{\Large $\tan \beta$}}
\end{minipage}
 \caption{Shown here are contours of the $\mu$-parameter in GeV in the  $\tan\beta$ versus $m_s$ (or equivalently $m_{3/2}$) plane. The blue shaded is excluded by the LEP bound on the chargino, the red region is excluded by BBN and in the green region gauge coupling unification is lost. The solid and dashed lines respectively indicate the band in which $m_h=125$ GeV can be achieved using susyHD. To make this plot we set $\zeta=1$; varying this parameter in a reasonable range has a small impact on this result. For $\zeta = 1$, and $m_s = 2\times 10^5 \text{ GeV}$, the corresponding gluino mass would be $M_3 \simeq 2000 \text{ GeV}$.  Up to small RGE corrections, the gluino mass can then be inferred for any other point in the plot using \eref{eq:MiToMf}, specifically $M_i \propto \zeta^2 m_s$.} 
   \label{fig:ParamSpace}
\end{figure}

In Fig.~\ref{fig:ParamSpace} we show contours of $\mu$ as a function of $m_s$ and $\tan\beta$. Here we choose $\zeta=1$ for concreteness, however we stress once again that the Higgsino mass is in practice independent of this choice. The dashed lines indicate $m_h$ as obtained with the susyHD code \cite{Vega:2015fna}, where we allow for the range 123 GeV $<m_h<127$ GeV to approximately account for the theoretical uncertainties. This calculation was chosen since it is tailored to mini-split scenarios.  We check that this is qualitatively similar to the result by Giudice et.~al.~\cite{Giudice:2011cg}.

In the limit of zero error bars on the Higgs mass, our choice of $\tan \beta$ then uniquely determines the stop masses, up to theoretical uncertainties. This in turn fixes the effective supersymmetry breaking scale $F/M$, and thus the rest of the scalar soft masses-squared.  Moreover the $b_{\mu}$ term and $a$-terms (which are suppressed, giving them negligible impact as discussed above) are fixed by requiring viable electroweak symmetry breaking given the constraints imposed in Eqs.~(\ref{eq:mu}) through (\ref{eq:aTerm}). 

Finally we discuss a number of additional constraints on the parameter space.  For high $m_s$, the Higgsino is heavier than 100 TeV, which is too heavy to allow for satisfactory gauge coupling unification \cite{Arvanitaki:2012ps}.  This is denoted by the green shaded region in the figure.  On the low $m_s$ end of the spectrum, there is the bound from LEP2 on the mass of the lightest chargino~\cite{Abdallah:2003xe}, as indicated by the blue shaded region. For some choices of $\zeta$, LHC constraints on the wino and/or gluino can also be relevant. We will discuss those in the next section. 

Finally, since the gravitino is the NLSP, there are important constraints from cosmology. The most robust constraint results from the need for late decays of the NLSP to the gravitino plus a $\gamma,\,Z^0,\,\mbox{or } h$ to not spoil big bang nucleosynthesis. If the NLSP is Higgsino-like, there will be a significant branching ratio to hadronic final states which could dissociate light nuclei. As a conservative bound, we therefore require the lifetime of the NLSP to be less than one second.  This will ensure that the NLSP has decayed before the start of BBN. (If the NLSP primarily decays into photons, this bound may be relaxed significantly.) The resulting constraint is shown by the red shaded region in the upper left corner of Fig.~\ref{fig:ParamSpace}. 

Moreover the requirement that the gravitino does not overclose the Universe places a strong upper bound on the reheating temperature after inflation~\cite{Moroi:1993mb, deGouvea:1997afu}. In particular, the Universe must reheat below the scalar masses, and even in this case the gravitino abundance places a strong constraint on the gluino mass \cite{Hall:2013uga}.  If the reheating temperature is above the gluino mass but below the scalar masses, the gluino should be $\lesssim 10$ TeV.  Given that 10 TeV gluino masses is well within the reach of a future 100 TeV collider~\cite{Cohen:2013xda, Beauchesne:2015jra, Arkani-Hamed:2015vfh}, this model results in interesting prospects for the LHC and beyond. 

\section{Collider Implications}\label{sec:ColliderPheno}

In this section we will briefly discuss some implications for the LHC and/or future colliders.  As we demonstrated in \S\,\ref{sec:Spectrum}, gauginos, and in particular the gluino, should be light enough to be observable for a wide region of parameter space. Since the messengers couple in an $SU(5)$ invariant way to \SB\!, the gaugino masses unify, implying
\begin{equation}
M_1: M_2:M_3 \simeq 1:2:6
\end{equation}
at the weak scale.  This can of course be relaxed easily in more general models, see for instance \cite{Cheung:2007es,Meade:2008wd,Buican:2008ws}. However, if we take these ratios seriously, it follows that an accessible gluino implies an accessible wino and bino. The Higgsino may also be light, although this is not necessary, see for instance second benchmark point in Table \ref{tab:ExampleParams}. In our simple model the NLSP is therefore always a bino or Higgsino-like neutralino. 

Since the gravitino LSP is heavier than $\sim10$ MeV, the NLSP is stable on detector length scales. The traditional chargino searches for pairs of $Z^0/W^\pm/h$ + missing energy provide the strongest constraints on this part of the spectrum. If the production cross section is primarily supplied by direct production of the chargino/neutralino pair, the bound on the wino is currently in the 400-450 GeV range~\cite{Aad:2015eda}.  

If the gluino is accessible, it will decay to the wino, bino or Higgsino through an off-shell squark. The lifetime is approximately given by
\begin{equation}
c\,\tau \sim 1\;\mathrm{cm} \left(\frac{10^6\;\mathrm{GeV}}{m_{\tilde q}}\right)^4 \left(\frac{M_3}{10^3\;\mathrm{GeV}}\right)^5,
\end{equation}
where~\cite{Gambino:2005eh} provides a careful calculation including radiative effects.  An LHC-accessible gluino then typically decays promptly for $m_{\tilde q}\lesssim 10^{6}$ GeV, but for the larger values of $m_s$ in Fig.~\ref{fig:ParamSpace}, the gluino could decay displaced or even be stopped in the detector. The latter two cases can yield an especially spectacular, although experimentally challenging signature.\footnote{See \cite{Liu:2015bma} for a recent reinterpretation of the CMS displaced dijet search \cite{CMS:2014wda} in terms of the mini-split parameter space.} 

On the other hand, if the decay is prompt, the gluino branching ratios are sensitive to the flavor texture of the squark mass matrices. Optimistically, it should be feasible to distinguish gauge mediation from other, flavor-generic versions of mini-split.  Concretely, in the absence of RGE effects, gauge mediation dictates that all the squarks are degenerate; the identity of $q$ would be democratic between all flavors of quark.  However, since there are several orders of magnitude separating the messenger scale from $m_s$, RGE effects on the squark spectrum need to be considered. As a result, the stop masses are subject to a 10-20\% suppression with respect to the soft masses of the other squarks.  This results in a tree-level branching ratio to top and bottom quarks between roughly 40\% and 50\% although these branching ratios can slightly shift once radiative corrections are accounted for \cite{Gambino:2005eh,Sato:2013bta}. Either way, an order-one fraction of gluino pair production events will yield four tops and missing energy, and should be relatively straightforward to discover at colliders. Moreover the branching ratio to tops and bottoms will further increase if the messenger scale is increased. However in this case the gravitino mass also increases, which strengthens the BBN bound in Fig.~\ref{fig:ParamSpace}. As this implies higher values of $m_s$, the gluino is then more likely to manifest displaced decays.  

Clearly if this model were realized in nature, a rich collider program would unfold at the 13 TeV LHC and/or a future proton machine.  In particular, the entire parameter space that is consistent with cosmological constraints with $m_{\tilde{g}} \lesssim 10 \text{ TeV}$ (when the reheat temperature is above the gluino mass) should be probeable with the data from a 100 TeV collider~\cite{Cohen:2013xda, Beauchesne:2015jra, Arkani-Hamed:2015vfh}.

\section{Conclusions}\label{sec:Conclusions}
In the paper we have introduced a model of gauge mediation with gaugino masses that are parametrically a loop factor below the scalar masses as the result of collective symmetry breaking.  This is relevant for models of mini-split supersymmetry that reproduce the observed Standard Model-like Higgs boson mass of 125 GeV while leading to gauginos that are observable at the LHC and/or a future 100 TeV collider.  Since it relies on gauge interactions to mediate \SB to the MSSM, this model does not suffer any of the flavor problems of the gravity + anomaly mediated models which have been rejuvenated in light of the Higgs boson discovery.  

We explored the parameter space of this example model and demonstrated that it is possible to achieve viable electroweak symmetry breaking, observable gauginos, and $\mu$ ranging from hundreds of GeV to many TeV.  We then discussed some rough expectations for the collider signatures.

Many models of gauge mediation can have suppressed gaugino masses.  However, these models tend to suffer from the $\mu-b_\mu$ problem.  Furthermore, while these models can in principle accommodate gauginos that are observable at the LHC and/or future colliders, they do not parametrically favor this scenario.  On the other hand, the model presented here accommodates a fully calculable and simple Higgs sector (given one fine-tuning to reproduce $m_Z$) and results in observable predictions for a wide range of parameters.  In the event that a gluino is discovered at the LHC (but the squarks are nowhere to be seen), correlating careful measurements of the branching ratios and possible displacement with the signature space of mini-split models will lead to deeper insights into the nature of \SB and its mediation. 

\section*{Acknowledgements}
We thank Csaba Csaki, Keiseke Harigaya, Markus Luty, Diego Redigolo, Giovanni Villadoro for useful discussions.  TC is supported by an LHC Theory Initiative Postdoctoral Fellowship, under the National Science Foundation grant PHY-0969510.
NC is supported by the Department of Energy under the grant DE-SC0014129. The work of SK is supported in part by the LDRD program of LBNL under under DoE contract DE-AC02-05CH11231.  We acknowledge the hospitality of the Aspen Center for Physics, supported by the National Science Foundation Grant number PHY-1066293, where parts of this work were completed.

\newpage

\bibliography{GMSBMiniSplit}
\bibliographystyle{utphys}
\end{document}